%% file: main.tex
\documentclass[%
 reprint,
superscriptaddress,
 amsmath,amssymb,
 aps,
]{revtex4-1}

\usepackage{graphicx}% Include figure files
\usepackage{dcolumn}% Align table columns on decimal point
\usepackage{amsmath}
\usepackage{bm}% bold math
\usepackage{siunitx}
\usepackage{float}
\usepackage{upgreek}
\usepackage{xcolor}
\begin{document}

\preprint{APS/123-QED}

\title{Acoustic spectral hole-burning in a two-level system ensemble}
\author{Andersson, G.}
\email{gandersson@uchicago.edu}
\affiliation{%
Department of Microtechnology and Nanoscience MC2, Chalmers University of Technology, SE-41296 G\"oteborg, Sweden
}
\author{Bilobran, A. L. O.}
\affiliation{%
Materials Science Institute (ICMUV), University of Valencia, E-46071 Valencia, Spain
}
\author{Scigliuzzo, M.}
\affiliation{%
Department of Microtechnology and Nanoscience MC2, Chalmers University of Technology, SE-41296 G\"oteborg, Sweden
}
\author{de Lima, M. M.}
\affiliation{%
Materials Science Institute (ICMUV), University of Valencia, E-46071 Valencia, Spain
}
\author{Cole, J. H.}
\affiliation{%
Chemical and Quantum Physics, School of Science, RMIT University, Melbourne VIC 3001, Australia
}
\author{Delsing, P.}
\affiliation{%
Department of Microtechnology and Nanoscience MC2, Chalmers University of Technology, SE-41296 G\"oteborg, Sweden
}
%\date{September 2020}

\begin{abstract}
 Microscopic two-level system (TLS) defects at dielectric surfaces and interfaces are among the dominant sources of loss in superconducting quantum circuits, and their properties have been extensively probed using superconducting resonators and qubits. We report on spectroscopy of TLSs coupling to the strain field in a surface acoustic wave (SAW) resonator. The narrow free spectral range of the resonator allows for two-tone spectroscopy where a strong pump is applied at one resonance while a weak signal is used to probe a different mode. We map the spectral hole burnt by the pump tone as a function of frequency and extract parameters of the TLS ensemble. Our results suggest that detuned acoustic pumping can be used to enhance the coherence of superconducting devices by saturating TLSs.
\end{abstract}

\maketitle
\subsection*{Introduction}
Two-level systems (TLSs) have attracted substantial interest in recent years \cite{Paladino2014,Mueller2019} as they are among the most important sources of loss limiting the performance of superconducting quantum circuits \cite{Martinis2005,Gao2008,Wisbey2010, Klimov2018,Burnett2019,Schloer2019}. Considerable efforts have been invested in attempting to understand their properties in order to mitigate this problem \cite{Burnett2014, DeGraaf2018}, and the dramatic improvements in the coherence of superconducting devices over the last decade are due mainly to design and process developments that reduce TLS loss. TLSs couple both to electromagnetic fields \cite{Gao2008, Wisbey2010, Pappas2011} and to strain \cite{Grabovskij2012, Anghel2007, Rogge1997} and recent approaches to investigating their behaviour typically involve measuring the lifetimes and resonance frequencies of superconducting resonators and qubits \cite{Shalibo2010,Klimov2018,Burnett2019,Schloer2019}. Much of the standard theory of two-level systems, however, was developed prior to the emergence of superconducting qubits, in experimental and theoretical research on their ultrasonic properties \cite{Phillips1987, Hunklinger1976}. In the present work we combine these approaches to probe TLSs in an acoustic resonator operating in the quantum regime where $\hbar \omega \ll  k_B T$.

\textcolor{black}{From experiments applying strain to superconducting qubit devices, there is evidence that the same individual TLSs couple to both strain and electric field \cite{Grabovskij2012, Lisenfeld2015, Lisenfeld2019}. Most TLSs inducing loss in superconducting devices are weakly coupled, and relaxation by phonon emission is generally understood to limit their lifetime \cite{Jaeckle1972, Mueller2019}. The strain associated with applied acoustic fields therefore should probe the same ensemble as circuit electric fields.}

Two-tone spectroscopy of TLSs has been performed using the fundamental and harmonic modes of a superconducting coplanar waveguide resonator \cite{Sage2011}, as well as more recently with the two normal modes of a system of coupled resonators \cite{Kirsh2017}. The large mode spacing of coplanar waveguide resonators limit these experiments to a small number of frequency points. In contrast, the short wavelength of sound \textcolor{black}{compared to electromagnetic radiation} gives rise to a spectrum with multiple closely-spaced modes with small differences in spatial distribution. This allows for two-tone spectroscopy where the pump-probe detuning is variable with a resolution corresponding to the mode spacing. By varying this detuning, it is possible to probe the shape of the spectral hole burnt in the TLS ensemble by the pump as a function of both frequency and power.

Surface acoustic wave (SAW) devices have been used in a number of quantum acoustic experiments, coupling mechanical modes to superconducting qubits \cite{Gustafsson2014}. Exotic regimes of atom-field interaction have been demonstrated \cite{Moores2017}, as well as the controlled generation of quantum states of SAW \cite{Satzinger2018}. Two-level systems have also been demonstrated to induce significant loss in SAW resonators at cryogenic temperatures \cite{Manenti2016}. Here, we exploit this effect as well as the narrow free spectral range of a SAW resonator to perform pump-probe spectroscopy of the TLS ensemble. 

\subsection*{Results}
\begin{figure}
	\centering
	\includegraphics[scale=0.3]{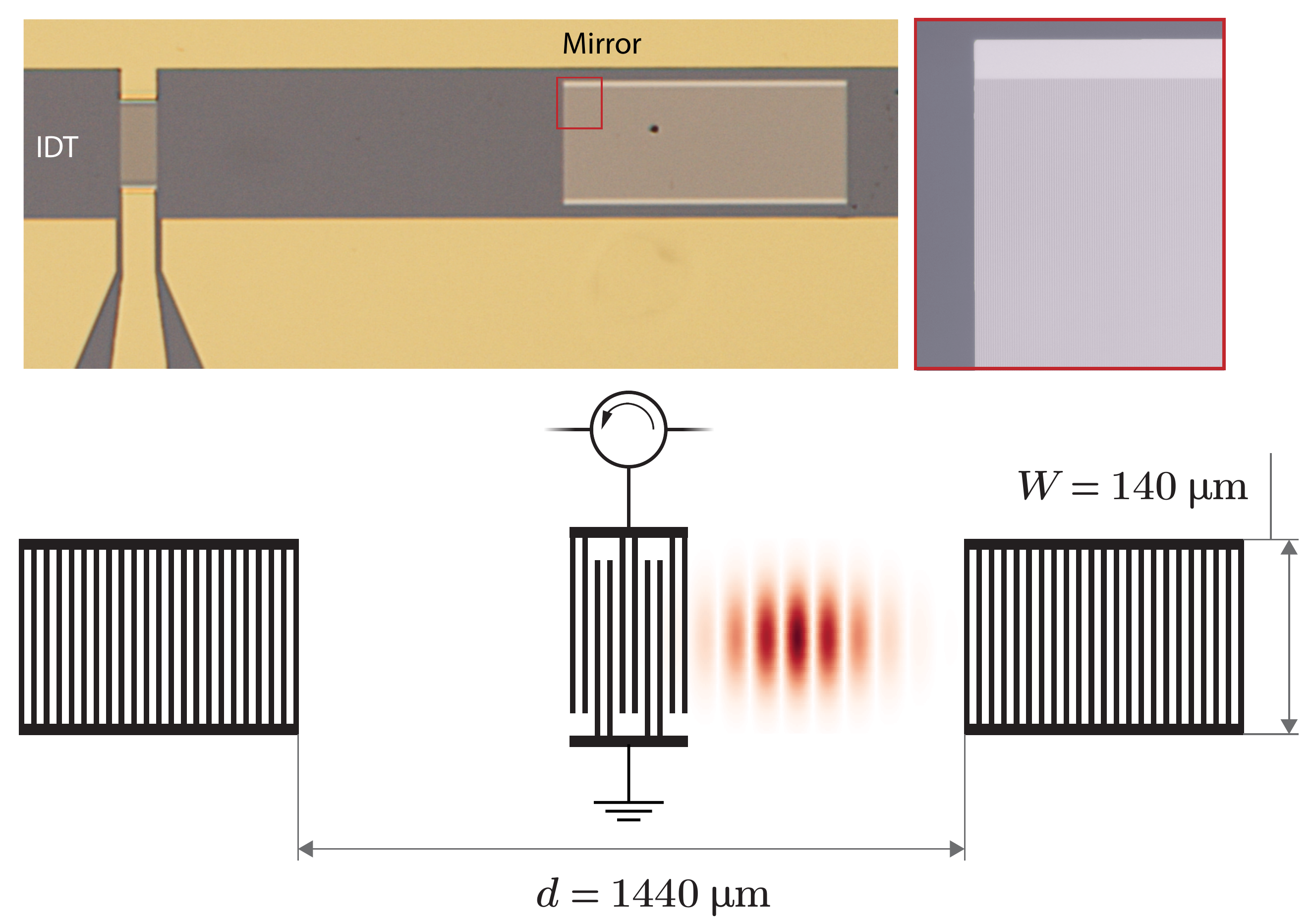}
	\caption{Device layout. \textcolor{black}{Microscope image showing the IDT and right hand Bragg mirror and schematic illustration of the SAW resonator. The IDT and mirrors are fabricated in aluminum on a GaAs substrate with gold ground planes. Inset shows part of the Bragg mirror.} The IDT with 50 periods provides a coupling port and is centered with respect to the Bragg mirrors. The mirrors have $N=800$ fingers each that are shorted together. As the length of the resonator is more than 1000 wavelengths, the spatial distributions of the different modes are nearly identical. The resonator is measured in reflection using a microwave circulator.}
	\label{fig: schematic}
\end{figure}

\begin{figure}
	\includegraphics[scale=0.75]{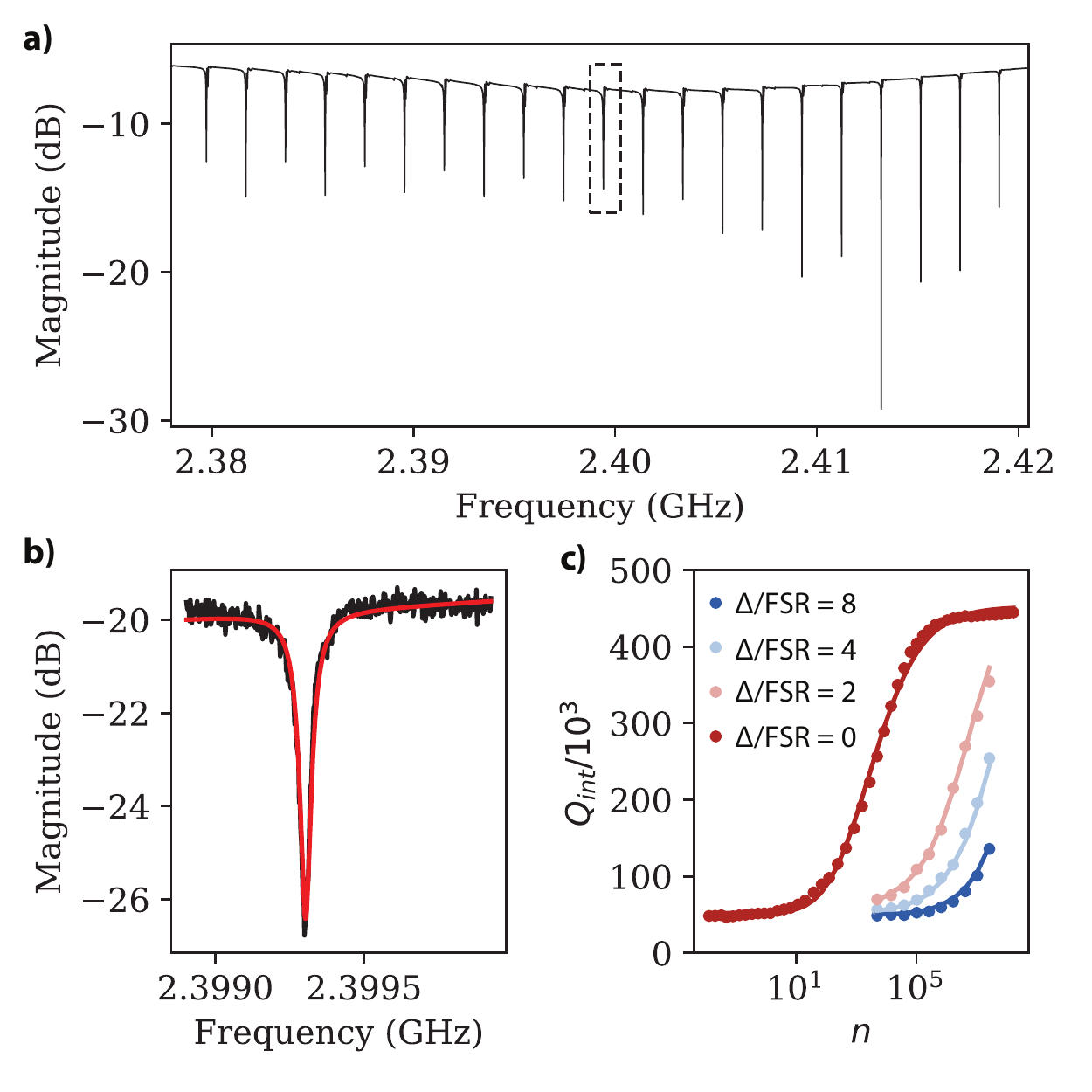}
	\caption{Resonator characterization. \textbf{a)} Magnitude of the reflection coefficient of the SAW resonator measured at high power ($n\approx 4\cdot 10^5$). The dashed outline indicates the pump mode at \SI{2.399}{GHz}. \textbf{b)} Example of fit to the pump mode resonance at low power ($n<1$). \textbf{c)} Internal Q-factor as a function of phonon number in the pump mode for the pump mode (red dots) as well as three probe modes below the pump. The pump-probe detunings are indicated and fits are shown as solid lines. The single mode loss saturation is fitted with eq.~\ref{eqQ}. The loss saturation in the detuned probe modes is fitted with eq.~\ref{eq: twotone loss}.}
	\label{fig: check}
\end{figure}
\subsubsection*{Single mode loss saturation}
While the microscopic nature of TLSs is still not well understood, the phenomenological standard tunneling model (STM) \cite{Machlup1954, Dutta1981, Mueller2019} describes successfully many of the low-temperature properties of amorphous solids. The STM models a TLS as a particle in a double-well potential, where two minima of similar energy are separated by a tunnel barrier. \textcolor{black}{Electric and strain fields deform the potential, inducing tunneling between the two states.} For the loss saturation of a single mode of frequency $f_r$, the standard theory derived from the STM gives \cite{Pappas2011,Mueller2019}
\begin{equation}
    \frac{1}{Q_\mathrm{int}(n)}=\frac{1}{Q_\mathrm{TLS}(0)}\frac{\tanh\frac{\hbar\omega_r}{2k_B T}}{\sqrt{1+\left(\frac{n}{n_C}\right)^\beta}}+\frac{1}{Q_\mathrm{res}},
    \label{eqQ}
\end{equation}
where $Q_\mathrm{int}$ is the internal Q-factor, $Q_\mathrm{TLS}$ the Q-factor corresponding to TLS loss and $n$ ($n_C$) is the average (critical) phonon number in the resonator. Residual internal loss not due to TLS is represented by $Q_\mathrm{res}$. The shape of the saturation curve is characterised by the phenomenological parameter $\beta$, where the scaling $\beta=1$ is expected from the standard theory, while values $\beta<1$ are commonly found in superconducting resonators. Since the temperature is low we use the approximation $\tanh{\left[\hbar\omega_r/(2k_B T)\right]}=1$ for the purpose of fitting. 

Figure 2 shows the resonator spectrum measured with a vector network analyzer. The internal Q-factor is extracted from fits to the data and plotted as a function of probe power is in Fig.~2c. We observe an order-of-magnitude change with power, indicating that SAW propagation losses at low temperature are limited by TLS loss. Fitting to eq.~(\ref{eqQ}), we find good agreement with theory and obtain $\beta = 1.05$, close to $\beta = 1$ as predicted by the STM.
\subsubsection*{Two-tone spectroscopy}
We now turn to the two-tone spectroscopy scheme. In this setup, a drive tone is applied at a mode in the center of the stopband ($\omega_r/2\pi= \SI{2.399}{GHz}$). As the power in the pump mode is swept, we measure the change in Q-factor and resonance frequency of all other modes \textcolor{black}{using a weak probe signal}. This yields the loss and frequency shift as a function of phonon number in the pump mode $n$ as well as pump-probe detuning $\Delta = \omega_\mathrm{probe}- \omega_\mathrm{pump}$. Due to the frequency dependence of the IDT response, all the modes have slightly different external quality factors. For this reason, the pump is fixed to a single mode to avoid the pump power adjustments necessary to get the same average phonon number in different modes. Based on the STM, expressions for the probe mode response were derived in \cite{Kirsh2017}, and give for the frequency shift
\begin{equation}
    \frac{\Delta \omega_r}{\omega_r \tanh\frac{\hbar\omega_r}{2k_B T}} = -\frac{3\sqrt{2}\tan{\delta}}{8}\frac{\Delta }{\Omega}\frac{\sqrt{1+\frac{\Omega^2}{2\Delta^2}}-1}{\sqrt{1+\frac{\Omega^2}{2\Delta^2}}+1}.
    \label{eq: twotone freqshift}
\end{equation}
The effective \textcolor{black}{Rabi frequency $\Omega$} of the pump may be expressed as $\hbar \Omega=2\gamma A_\mathrm{pump}$, where $\gamma$ is the average elastic dipole moment coupling the TLS to the pump field of amplitude $A_\mathrm{pump}$. This implies the drive strength depends on the average phonon number in the resonator as $\Omega \propto A_\mathrm{pump}\propto \sqrt{n}$. Here, $\tan{\delta}$ is the dielectric loss tangent due to TLS. The frequency shift moves probe modes closer to the pump mode and for a given $\Delta$ it is maximized by a finite $n=n_{max}$. This can be understood from the dispersive interaction, where each off-resonant ground state TLS contributes a shift $\Delta\omega_{r,i}=g_i^2/(\omega_r-\omega_i)$. Saturating TLSs disables this interaction. For pump phonon numbers $n>n_{max}$, TLSs on both sides of the probe mode get saturated, and the frequency shift starts to diminish. For $n<n_{max}$ TLSs are saturated predominantly on one side of the probe mode.

\textcolor{black}{The same model also yields the loss in two-tone spectroscopy}. The change in probe mode loss is symmetric around the pump and given by \cite{Kirsh2017}
\begin{multline}
     \delta \left( \frac{1}{Q_\mathrm{TLS}}\right) \frac{Q_\mathrm{TLS}}{\tanh\frac{\hbar\omega_r}{2k_B T}} = \\- 1- \left( \frac{\Delta}{\Omega}\right)^2\left[6 +3X 
     \ln\left(1+\left( \frac{\Omega}{\Delta}\right)^2 \left(1-X\right)\right)\right],
     \label{eq: twotone loss}
\end{multline}
where
\begin{equation}
    X = \sqrt{1+2\left( \frac{\Delta}{\Omega}\right)^2}.
\end{equation}
The internal Q-factor as function of pump power of modes detuned from the pump by 2, 4 and 8 times the FSR, respectively, is shown in Fig.~2c along with fits to eq.~(\ref{eq: twotone loss}). Comparing to the single mode loss, we note that saturation in detuned probe modes occurs at higher pump powers. \textcolor{black}{In the single mode measurement, TLS losses are reduced by more than 90 \% at pump phonon numbers of $n=n_s\approx 10^5$. At this power, still only TLSs near-resonant with the pump mode are saturated, with little impact on the loss in other modes}. As the pump power is increased further, TLSs are saturated across a wider frequency span, reducing losses in nearby modes. This is illustrated in Fig~3a.
%\begin{figure}
%	\centering
%	\includegraphics[scale=0.75]{figures/pumpprobe_graphic2.pdf}
%	\caption{Illustration of TLS saturation due to the pump. Near-resonant TLSs in the ground state can be excited by the SAW field and induce loss. Saturated TLSs do not contribute to loss. At the single mode saturation phonon number $n_s$, TLS loss is mitigated in the pump mode, while losses in other modes are only marginally affected. As the strength of the pump tone is increased, the spectrum of the saturated TLSs becomes wider, reducing loss substantially in nearby modes.}
%	\label{fig: twotone_schematic}
%\end{figure}

The full probe response as a function of detuning and pump power is shown in Fig.~3. The loss due to TLS is plotted in Fig.~3a and a fit to eq.~(\ref{eq: twotone loss}) is shown in Fig.~3c. For pump phonon numbers $n>n_s$, the losses in the pumped mode are completely saturated and no additional effects of increased pump power can be resolved. The shape of the spectral hole, however, depends strongly on power in the entire range accessible in this experiment. The spectral hole due to the pump continues to widen as the pump power is increased even though pump mode losses are completely saturated. Conversely, holes in the absorption spectrum burnt using pump phonon numbers below $n_s$ are not well resolved by the free spectral range of the resonator. Figure~3b shows the measured frequency shifts, and the fit to eq.~\ref{eq: twotone freqshift} is plotted in Fig.~3d. We observe that unlike the loss saturation, the frequency shift is not monotonic in power for a given detuning $\Delta$. This is consistent with our model of the dispersive interaction between off-resonant TLSs and the SAW modes.
\subsubsection*{Strength of SAW-TLS interaction}
In the single mode measurements, we observe a scaling of the effective Rabi frequency with phonon number that is consistent with the STM. We extract $\beta=1.05$ which is in good agreement with $\Omega \propto \sqrt{n}$. However, in the two-tone experiment the behaviour is different. The scaling of the extracted effective Rabi frequency characterising the coupling of the modes to the TLS bath does not correspond to the expected $\Omega \propto \sqrt{n}$. Instead, we find $\Omega \propto n^k$ with $k\approx 0.3$, with a slight divergence between the values obtained from the frequency shift and loss. The Rabi frequency $\Omega$ as a function of pump strength is shown separately for the loss and frequency shift fits in Fig.~4. A mechanism that has been suggested to cause slower scaling than square root in the loss saturation with power is based on TLS-TLS interactions that cause TLSs to fluctuate in frequency in and out of resonance with the probe mode \cite{Faoro2012, Burnett2014, Faoro2015}. There are two main differences between the single mode and two-tone experiments. In addition to the pump being off-resonant in the two-tone case, the power is also substantially higher ($n>n_s$). It is not clear why TLS-TLS interactions should play a larger role in the two-tone experiment. One reason could be that the effect of TLSs drifting in frequency between the pump and probe frequencies is more important in this case.

\textcolor{black}{A simpler model for the two-tone spectroscopy was derived in \cite{Capelle2020}. Under the assumption of uniform coupling strength of TLSs to the resonant modes, the spectral hole has a Lorentzian lineshape. In Supplementary note 1, we repeat the analysis of our two-tone spectroscopy data using this model and obtain a scaling for the Rabi frequency of $\Omega \propto n^{0.3}$ (Supplementary Figs.~1-3). That a similar scaling is obtained with different models suggests the disparate scaling between on- and off-resonant pumping is a real effect, although it does not explain its physical mechanism.}

If we instead fit eqs.~(\ref{eq: twotone freqshift}, \ref{eq: twotone loss}) with the power as independent variable, \emph{assuming} $\Omega = \Omega_0\sqrt{n}$ for each probe mode, we can extract an average single phonon Rabi frequency $\Omega_0/2\pi = \SI{25}{kHz}$. This is of the same order as what is found in \cite{Kirsh2017}. In order to obtain an estimate for the intrinsic TLS linewidth, we rewrite the denominator of the single mode power dependence given by eq.~\ref{eqQ} in terms of the effective Rabi frequency and loss rates as
\begin{equation}
    n/n_C = \Omega^2T_1T_2
\end{equation}
where $T_1$ and $T_2$ are the TLS excited state lifetime and coherence time, respectively. This implies that at above the critical phonon number $n_C$, the effective Rabi frequency exceeds the effective loss rate $\Omega>1/\sqrt{T_1 T_2}$. Still assuming the $\Omega = \Omega_0\sqrt{n}$ behaviour for the drive strength, and that $T_2= 2T_1$, we can extract an approximate characteristic TLS coherence time $T_2\approx \SI{2}{\upmu s}$. While obtaining this value relies on many assumptions, it appears consistent with the width of the spectral holes observed in the two-tone spectroscopy.

\begin{figure*}
	\centering
	\includegraphics[scale=0.75]{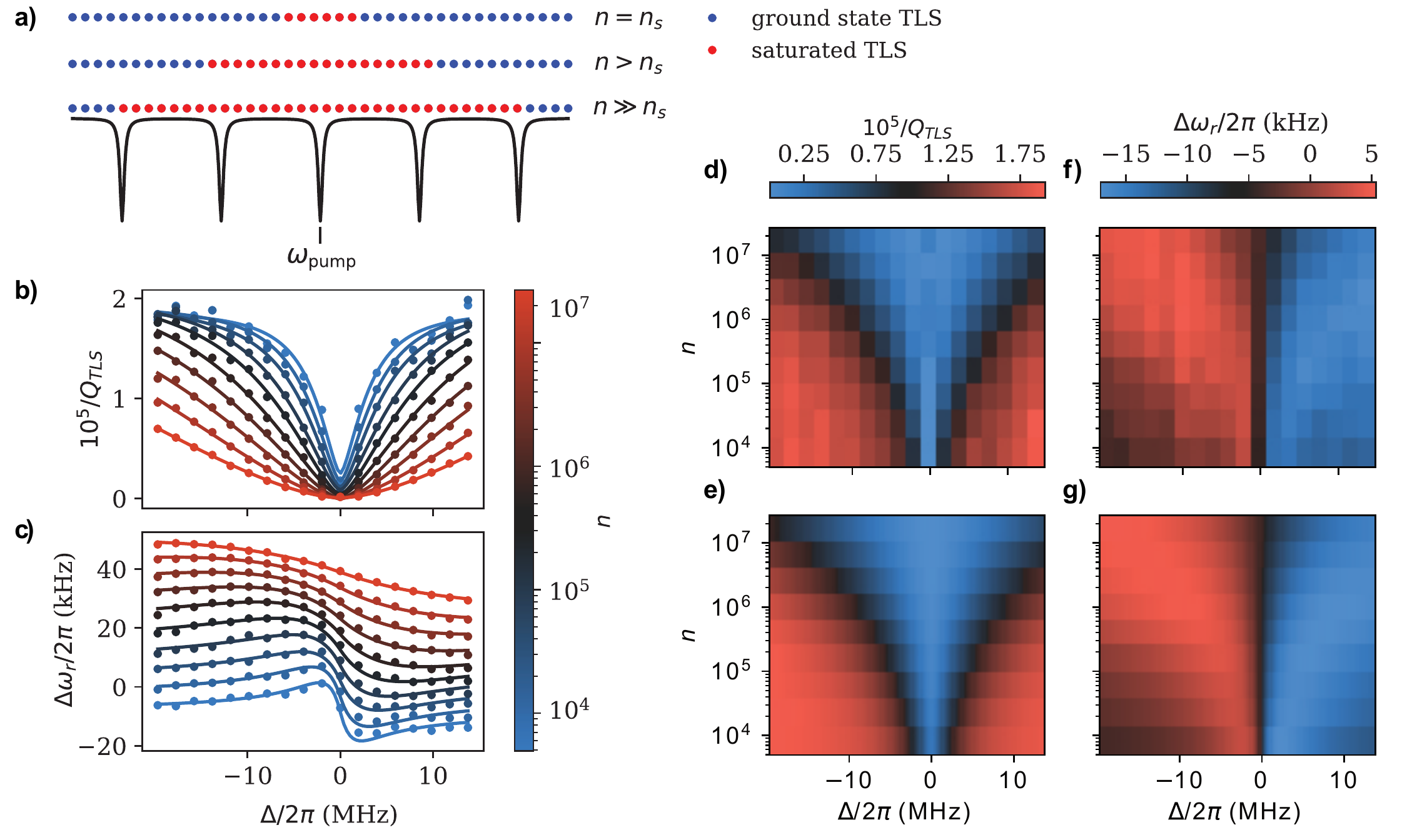}
	\caption{\textcolor{black}{Two-tone spectroscopy of the TLS ensemble. \textbf{a)} Illustration of TLS saturation due to the pump. Near-resonant TLSs in the ground state can be excited by the SAW field and induce loss. Saturated TLSs do not contribute to loss. At the single mode saturation phonon number $n_s$, TLS loss is mitigated in the pump mode, while losses in other modes are only marginally affected. As the strength of the pump tone is increased, the spectrum of the saturated TLSs becomes wider, reducing loss substantially in nearby modes. In two-tone spectroscopy, a drive tone is applied at the pump mode and response in the other modes are measured as a function of detuning $\Delta$ and the number of pump phonons $n$. The TLS loss $1/Q_\mathrm{TLS}$ is shown in \textbf{b)} with fits to eq.~\ref{eq: twotone loss} (solid lines). The color scale indicates the pump phonon number. Applying a detuned pump also induces a frequency shift in the probe mode, plotted in \textbf{c)} with fits to eq.~\ref{eq: twotone loss} (solid lines) and a \SI{5}{kHz} offset between traces. In \textbf{d)} we plot the TLS loss as a function of detuning and pump phonon number, with the fit shown in \textbf{e)}. The corresponding plots for the frequency shift are shown in \textbf{f)-g)}. The frequency resolution in $\Delta$ in the measurement is given by the free spectral range of the resonator.}}
	\label{fig: twotone_col}
\end{figure*}

\begin{figure}
	\centering
    \includegraphics[scale=0.75]{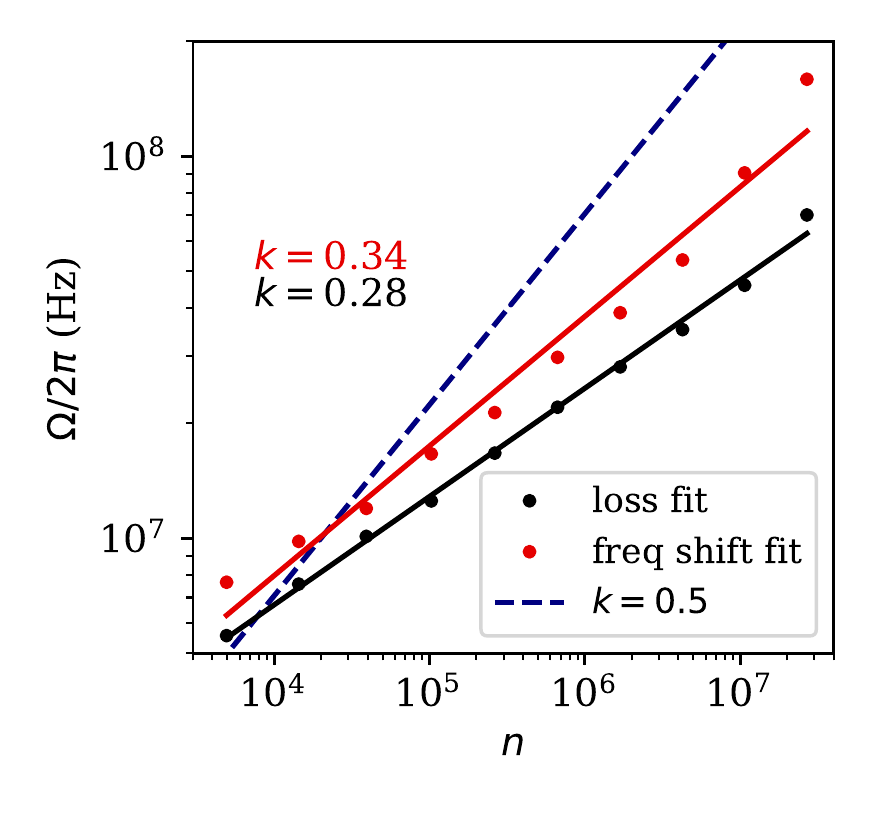}
	\caption{Effective Rabi frequency of the pump extracted from fits as function of number of pump phonons. Data shown in black derives from fits to the probe loss in two-tone spectroscopy (Fig.~\ref{fig: twotone_col}b). The frequency shift (Fig.~\ref{fig: twotone_col}c) yields smaller values for the pump strength at high phonon numbers, shown here in red. Solid lines show fits to extract the scaling. The blue line shows the $k=0.5$ slope predicted by the STM, corresponding to $\Omega \propto \sqrt{n}$. }
	\label{fig: twotone_loglog}
\end{figure}
\subsection*{Discussion}
Our measurements have revealed the shape of the spectral hole burnt in a TLS ensemble by a strong pump. We have shown that the response in acoustic susceptibility due to pumping is qualitatively well captured by theory based on the STM \cite{Kirsh2017}, but find a deviation in the scaling of the extracted \textcolor{black}{Rabi frequency} with pump power. From our measurements we extract estimations of the average single phonon Rabi frequency and linewidth of the TLS distribution. 

Our results suggest using acoustic pumping to mitigate TLS loss in superconducting qubits. While improvements in design and fabrication methods have led to a rapid increase in coherence in recent years, active means of TLS saturation \cite{Matityahu2019} are hard to develop due to the incompatibility of resonant microwave pump fields with device functionality. This limitation does not necessarily apply to the acoustic pumping scheme demonstrated here. For a spectral hole due to an applied SAW drive, it is straightforward to reach a linewidth of several tens of MHz, which allows for pumping at sufficient detuning to prevent spurious excitations due to crosstalk, but still within range of the spectral hole. \textcolor{black}{While high-coherence superconducting devices are not compatible with piezoelectric substrates \cite{Scigliuzzo_2020}, the generation of high-frequency SAW on non-piezoelectric substrates, including silicon, is well established \cite{Kino1973,Bueyuekkoese2013,Yuan2017}. Combining SAW and superconducting qubits could then be done by fabricating IDTs on local piezoelectric thin films while the qubits are placed directly on the non-piezoelectric substrate.} Acoustic resonators also have a lower intrinsic sensitivity to temperature than superconducting devices, which should make them well suited for studying the temperature dependence of TLS-induced noise and dissipation across a wide temperature range.

\subsection*{Methods}
Our device, shown schematically in Fig. 1, consists of two Bragg mirrors with 800 fingers each, separated by a distance $d=\SI{1440}{\upmu m}$. An interdigital transducer (IDT) at the center of the cavity provides an input and output port. The mirrors and IDT are fabricated from aluminium on a piezoelectric GaAs substrate. The IDT has 50 periods of $\SI{1.2}{\upmu m}$ and a split-finger design to suppress mechanical reflections \cite{Bristol1972}. The mirrors provide a stopband of approximately \SI{40}{MHz} centered around \SI{2.40}{GHz}. As the IDT is spatially centered with respect to the mirrors, only even modes can be excited, giving an effective free spectral range of $\textup{FSR}=v_\textup{SAW}/L=\SI{1.97}{MHz}$, supporting 20 modes within the stopband. Due to the finite penetration depth (approx. \SI{15}{\upmu m}) of the field into the reflectors, the effective cavity length $L$ is greater than the mirror edge separation $d$. To probe the resonator we measure the reflection off the IDT via a circulator. In two-tone spectroscopy the pump tone is also applied at the same port. The measurements presented here are performed at the \SI{10}{mK} base temperature of a dilution refrigerator.

\subsection*{Acknowledgment}
This work was supported by the Swedish Research Council, VR and by the Knut and Alice Wallenberg foundation. This project has also received funding from the European Union's Horizon 2020 research and innovation programme under grant agreement No 642688 (SAWtrain). JHC is supported by the Australian Research Council Centre of Excellence programme through Grant number CE170100026 and the Australian National Computational Infrastructure facility.

%\bibliography{citations_TLS}
\input{main.bbl}
%\bibliographystyle{naturemag}

\end{document}

\preprint{APS/123-QED}
\title{Supplementary materials for acoustic spectral hole-burning in a two-level system ensemble} 
%\date{\today}
\renewcommand{\figurename}{Supplementary}
\renewcommand{\thefigure}{Figure \arabic{figure}}
\renewcommand{\theequation}{S\arabic{equation}} 

\author{Andersson, G.}
\email{gustav.andersson@chalmers.se}
\affiliation{%
Department of Microtechnology and Nanoscience MC2, Chalmers University of Technology, SE-41296 G\"oteborg, Sweden
}
\author{Bilobran, A. L. O.}
\affiliation{%
Materials Science Institute (ICMUV), University of Valencia, E-46071 Valencia, Spain
}
\author{Scigliuzzo, M.}
\affiliation{%
Department of Microtechnology and Nanoscience MC2, Chalmers University of Technology, SE-41296 G\"oteborg, Sweden
}
\author{de Lima, M. M.}
\affiliation{%
Materials Science Institute (ICMUV), University of Valencia, E-46071 Valencia, Spain
}
\author{Cole, J. H.}
\affiliation{%
Chemical and Quantum Physics, School of Science, RMIT University, Melbourne VIC 3001, Australia
}
\author{Delsing, P.}
\affiliation{%
Department of Microtechnology and Nanoscience MC2, Chalmers University of Technology, SE-41296 G\"oteborg, Sweden
}

\maketitle
\section{Additional analysis of the spectral hole}
In the reference \cite{Capelle2020} a model for the loss and frequency shift in two-tone spectroscopy is introduced that assumes a uniform (average) coupling strength of TLSs to the resonant modes. For the frequency shift and loss, respectivley, they derive
\begin{equation}
    \Delta \omega_r = -\frac{\Gamma_0}{2}\frac{\left(\Delta/\Gamma_2\right) \Tilde{n}}{\sqrt{1+\Tilde{n}}\left[\left(\Delta/\Gamma_2\right)^2+\left(1\sqrt{1+\Tilde{n}}\right)^2\right] },
    \label{eq: fshift_CNRS}
\end{equation}
\begin{equation}
    \frac{\Gamma_{int}}{\Gamma_0} =1 -\frac{\Tilde{n}\left(1+\sqrt{1+\Tilde{n}}\right)}{\sqrt{1+\Tilde{n}}\left[\left(\Delta/\Gamma_2\right)^2+\left(1\sqrt{1+\Tilde{n}}\right)^2\right] }.
    \label{eq: loss_CNRS}
\end{equation}
Here, $\Tilde{n}=n/n_c$ is the average phonon number scaled with the critical phonon number. The maximal resonator damping due to TLS is set by $\Gamma_0$ and $\Gamma_2$ represents the intrinsic TLS linewidth. This is related to the effective Rabi frequency via \cite{Capelle2020}
\begin{equation}
    \Omega = \Gamma_2\sqrt{1+\Tilde{n}} \approx \Gamma_2\sqrt{\Tilde{n}}.
\end{equation}
Fits of our two-tone spectroscopy data to Eqs.~\ref{eq: fshift_CNRS}-\ref{eq: loss_CNRS} are shown in Supplementary Figs.~1-2. The average and critical phonon numbers are \emph{not} free parameters in this fit, but determined from the single mode loss saturation measurement (See Fig.~2 of the main text). This implies that the scaling of $\Omega$ with power will appear as a power dependence of the extracted $\Gamma_2$ value when traces are fit independently. Fits for different pump powers do not yield a uniform $\Gamma_2$, but a power dependent behavior shown in Supplementary Fig.~3. This power dependence of $\Gamma_2$ is approximately
\begin{equation}
    \Gamma_2 = \Gamma_{2,0}n^{-0.25}.
\end{equation}
Inserting this power dependence we get a scaling of the effective Rabi frequency with pump phonon number as
\begin{equation}
    \Omega \propto n^{0.25}.
\end{equation}
This behavior is similar to what we obtain in the analysis based on expressions derived in \cite{Kirsh2017}.
\begin{figure}
	\centering
	\includegraphics[scale=1]{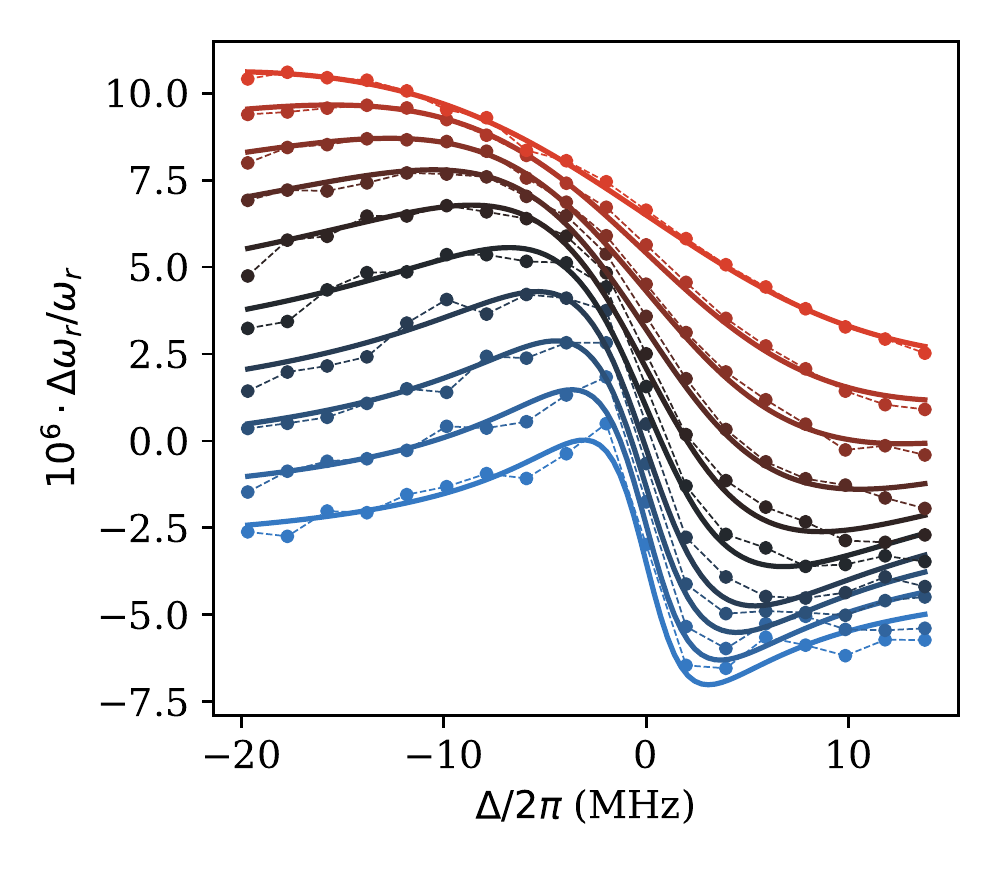}
	\caption{Relative frequency shift measured in two-tone spectroscopy with fits to Eq.~\ref{eq: fshift_CNRS}}
	\label{fig: fshift_CNRS}
\end{figure}

\begin{figure}
	\centering
	\includegraphics[scale=1]{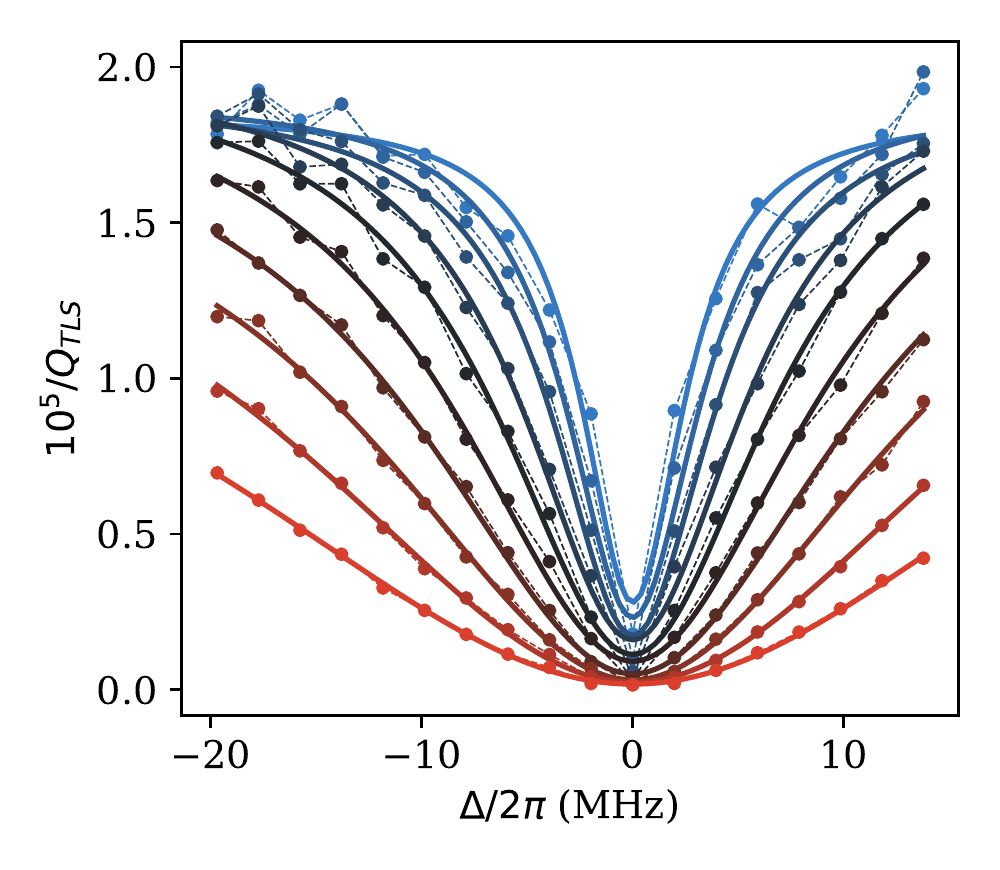}
	\caption{Loss rate measured in two-tone spectroscopy with fits to Eq.~\ref{eq: loss_CNRS}}
	\label{fig: loss_CNRS}
\end{figure}
\begin{figure}
	\centering
	\includegraphics[scale=1]{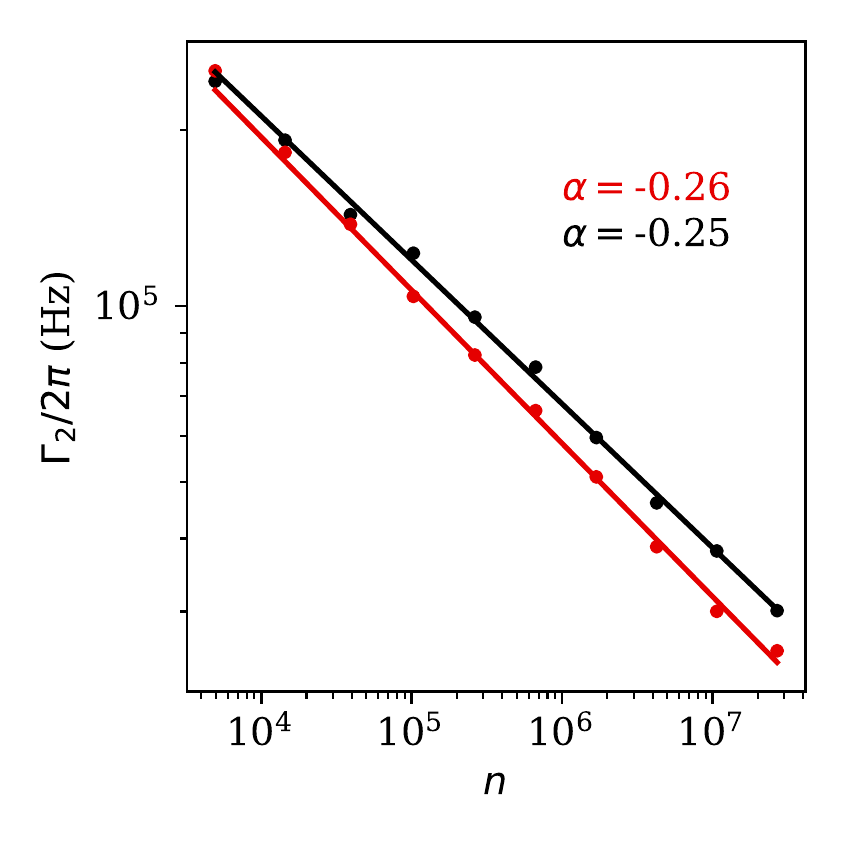}
	\caption{Intrinsic TLS linewidth extracted from two-tone spectroscopy measurements using fits to Eqs.~\ref{eq: fshift_CNRS}-\ref{eq: loss_CNRS}. The presence of a power dependence indicates a scaling of the effective Rabi frequency with phonon number slower than square-root. Values extracted for the loss (frequency shift) are shown in black (red). }
	\label{fig: loglog_CNRS}
\end{figure}